# Visualizing Engineering Fundamentals: Design of Mixed Reality and Physical Toolkits for Effective Learning

Mohammad Abu Nasir Rakib[1], Sharmin Akter[1], Eshwara Prasad Sridhar[1], Somik Biswas[1],

Md Rassel Raihan[1], Mahmudur Rahman[1]

[1]The University of Texas at Arlington, Arlington, Texas, United States

**Running head:** Human Factors in Engineering Learning Tool Design

**Corresponding Author:** Correspondence concerning this article should be addressed to

Mahmudur Rahman, Woolf Hall Room 420, 500 West First Street, Arlington, TX 76019.

Email: mrahman@uta.edu

**Feature at a Glance**

This study examined students' experiences with mixed-reality applications and physical toolkits in Engineering Mechanics to inform design guidelines for educational tools. In a user study with 24 participants, we compared classroom instruction alone, classroom instruction with a mixed-reality application, and classroom instruction with physical toolkits. Thematic analysis of participant feedback revealed that learners' workflows and engagement with fundamental mechanics problems varied across instructional modalities. Participants valued multimodal and interactive experiences that combined visualization with hands-on interaction, while reporting challenges with complex or unclear visualizations. These insights support the human-centered design of mixed-reality and physical tools for engineering education.

# Visualizing Engineering Fundamentals: Design of Mixed Reality and Physical Toolkits for Effective Learning

Virtual and mixed reality have become widely adopted technologies for training complex skills across a variety of domains. Prior research has demonstrated their benefits in architectural design (Ummihusna & Zairul, 2022), medical surgery (Ruthenbeck & Reynolds, 2015), art and drawing (Song & Li, 2018), industrial design (Bernardo & Duarte, 2021), and classroom instruction (Häfner et al., 2013; Syed et al., 2017). The growing appeal of these technologies stems from their comparatively low implementation cost, their suitability for gamified learning environments, and their potential to reduce unnecessary cognitive load while improving conceptual understanding. Despite these advantages, virtual and mixed reality continue to face notable limitations that affect accessibility and usability, especially in educational settings. Challenges commonly reported in the literature include insufficient user-centered design practices (Escallada et al., 2025), the need for specialized technical skills to create new interaction techniques or physics-based behaviors (Yigitbas et al., 2021), and the labor-intensive development of problem scenarios, which often leads to low-fidelity or narrowly scoped applications (Ashtari et al., 2020). Still, virtual and mixed reality remains an appealing tool for instruction, as its immersive qualities consistently support higher engagement, enjoyment, and motivation among learners (Justo et al., 2022).

## Virtual and Mixed Reality in Engineering Education

A substantial body of research has examined how virtual learning applications can support discipline-specific skill development, particularly through interaction techniques that enable learners to engage with spatial, mechanical, or procedural content. These applications are not simply to improve motivation to learn or reduce cognitive load, but to enable forms of interaction and visualization that are difficult or impossible to achieve with traditional teaching tools (Chi et al., 2025; Ekman et al., 2024; Häfner et al., 2013; Kamińska et al., 2017). These interactive experiences simulate real-world manipulation to support users' spatial reasoning and mechanical understanding, including rotating, pushing, pulling, grabbing, twisting, and screwing/unscrewing components to mirror physical handling tasks. For example, Gregg et al. (2024) developed a medium-fidelity VR assembly tool that enabled learners to rotate and align parts with contextual labels to clarify component relationships. Other work has extended these interaction strategies to support tool handling, internal machine adjustments (Michels & Häfner, 2022; Park et al., 2023), and material behavior exploration, including stress–strain visualization and failure modes (Perez & Keleş, 2025). Beyond assembly and manipulation, VR has been used to support 3D modeling through operations such as adding and subtracting geometry, pulling edges or vertices, duplicating shapes, and adjusting scale (Brás et al., 2020). Another study further introduced a deformation library that allows users to sculpt and bend virtual objects in ways that resemble clay manipulation (Luo et al., 2025). Although these approaches provide rich interaction possibilities, VR environments still lack the tactile cues available in hands-on work with physical parts. To address this, recent systems have incorporated haptic interfaces (Wee et al., 2021) and motion or eye-tracking technologies (Clay et al., 2019) to deliver real-time feedback and enhance action–perception coupling during learning tasks.

**Study Overview**

This study examines how immersive and tangible visualization approaches can support the learning of fundamental mechanical engineering concepts. We investigated learners' interactions with a mixed reality application and a set of physical 3D models representing the same core engineering problem (Figure 1), focusing on how visualization modality and model tangibility influence understanding of abstract mechanics concepts. Rather than evaluating VR effectiveness in general, the study emphasizes how different forms of visualization, immersive, interactive, and hands-on, shape conceptual reasoning. Participants completed baseline and post-study knowledge quizzes to assess learning gains and responded to open-ended questions that were analyzed thematically. This mixed-methods approach enabled the identification of learner expectations and concrete design considerations to inform the development of future interactive learning tools for foundational engineering education.

**Figure 1**

*Exploring potential challenges with MR and ET, and future design ideas in engineering instructional tool design.*

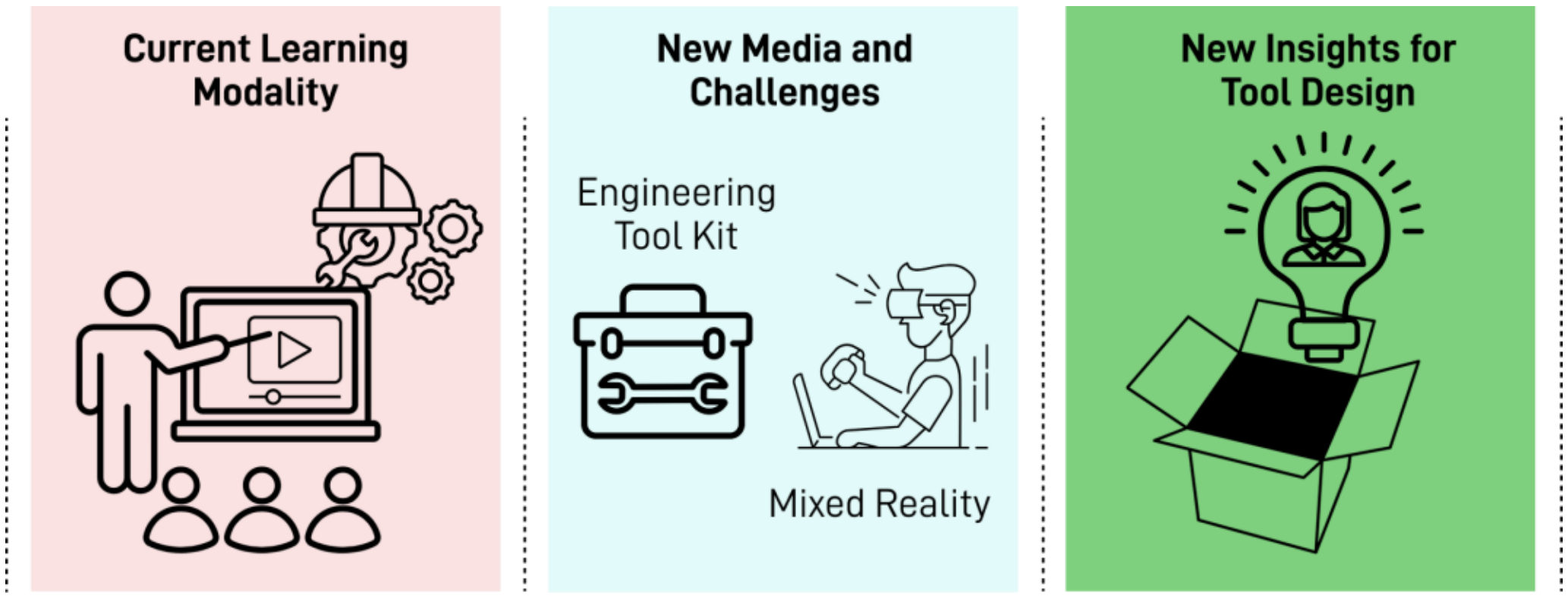

## Study Design

We conducted a 90-minute user study (Figure 2) to explore how participants interact with three instructional modalities: (a) the Traditional Classroom (TC) instruction alone, (b) TC instruction with a engineering toolkit (ET) (imitation model of beam, supports, and concentrated load for an example problem), and (c) TC instruction with a Mixed Reality (MR) application with a virtual model of the same example problem. The activities were designed to examine various aspects of problem-solving and engagement with fundamental engineering mechanics lessons, such as beam theory and beam bending behavior under different support and loading conditions. Twenty-four participants were recruited from the student population at the University of Texas at Arlington (UTA) who had completed a foundational mechanics (Engineering Statics) course. All study procedures were reviewed and approved by the UTA Institutional Review Board (UTA IRB No. 2025-0199). Participants received $20 as compensation for their time and contributions.

**Figure 2**

*An overview of the study procedure.*

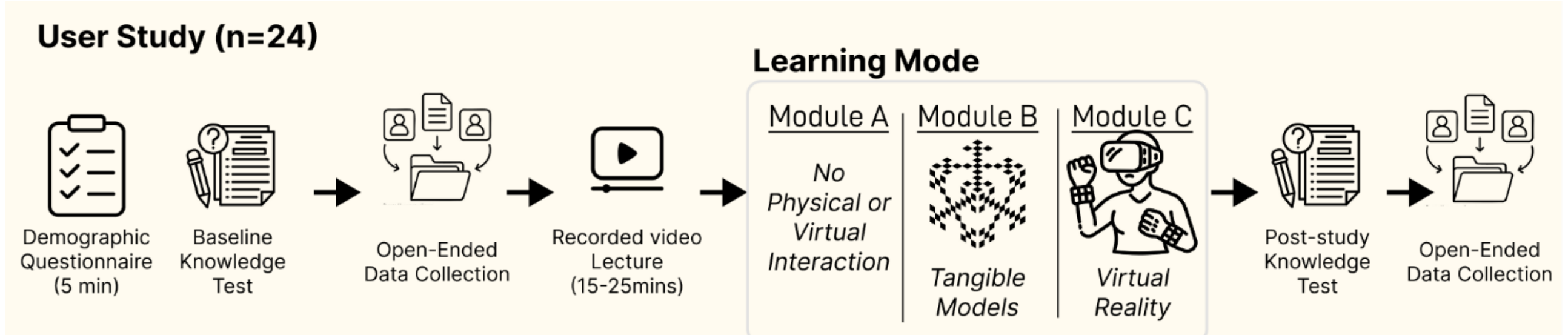

**Study Procedure**

After providing informed consent, participants completed a demographic survey, a baseline knowledge quiz, and three open-ended questions. The baseline quiz was administered to verify equivalence in prior knowledge across experimental groups. All participants then viewed a recorded lecture on beam bending under different loading and support conditions, including a worked example of a simply supported beam subjected to an external load. Participants were provided with a pen and paper to take notes while watching the lecture video, creating a classroom-like learning environment. Following the lecture, participants were assigned to one of three conditions. The traditional classroom (TC) group completed a knowledge quiz to assess learning outcomes. The experimental toolkit (ET) group received a physical toolkit and a handout with instructions and conducted hands-on experiments by adjusting the load location (Figure 3A). The mixed reality (MR) group interacted with an MR application featuring a 3D beam model with dynamically updated shear force and bending moment diagrams (Figure 3B). After completing the experiential activities, participants in the ET and MR groups completed the knowledge quiz. After the knowledge quiz, all participants were given a post-study survey containing additional open-ended questions. The pre- and post-lecture open-ended questions are given in Table 1. Participants' responses to these questions, combined with observations of their interactions, formed the basis for our qualitative data analysis.

**Figure 3**

*Participants interacting with toolkits (A) and an MR application(B).*

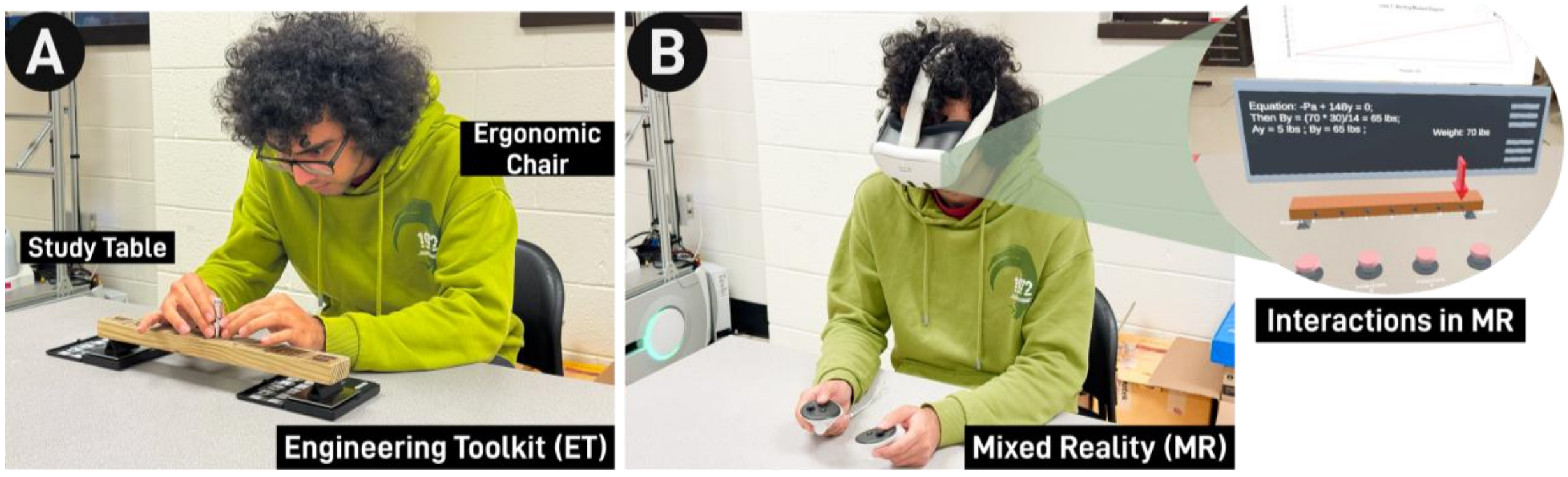

## Thematic Analysis and Theme Generation

Three researchers with expertise in computer science and design, mechanical engineering, and human factors conducted the thematic analysis (Braun & Clarke, 2021; Terry et al., 2017). Responses to the open-ended surveys (Table 1) were independently reviewed and coded, and relevant quotes and observations were documented. The coding process was refined iteratively until the researcher reached consensus. Axial coding was then applied to organize the initial codes into higher-level thematic clusters. We identified four key themes that emerged from empirical data (Figure 4). We also examined participants' pre- and post-study quiz results and compared them with their qualitative responses to identify any patterns or consistencies between performance and reported experiences.

**Table 1**

*Open-ended questions.*

| Pre-Lecture Open-ended Survey | Post-Lecture Open-Ended Survey |
|---|---|
| 1. What challenges did you face in answering the above questions?<br>2. What could help improve your comprehensive understanding of them?<br>3. How did you come up with the solutions to these questions? | 1. What challenges did you face in answering the above questions?<br>2. What helped you solve the questions?<br>3. How effective were the learning materials in helping you understand the concepts?<br>4. What could be added to the teaching methods (digital or physical materials) to improve the learning experience, and why? |

| Pre-Lecture Open-ended Survey | Post-Lecture Open-Ended Survey |
|---|---|
| | 5. How engaging or frustrating were the learning materials, and how could they be made more engaging for students? |

*Note*. In the pre-lecture survey, the phrase "the above questions" refers to the baseline knowledge quiz questions. On the other hand, in the post-lecture survey, the same phrase refers to the knowledge quiz that is based on the lecture.

**Figure 4**

*Mapping diagram for theme generation.*

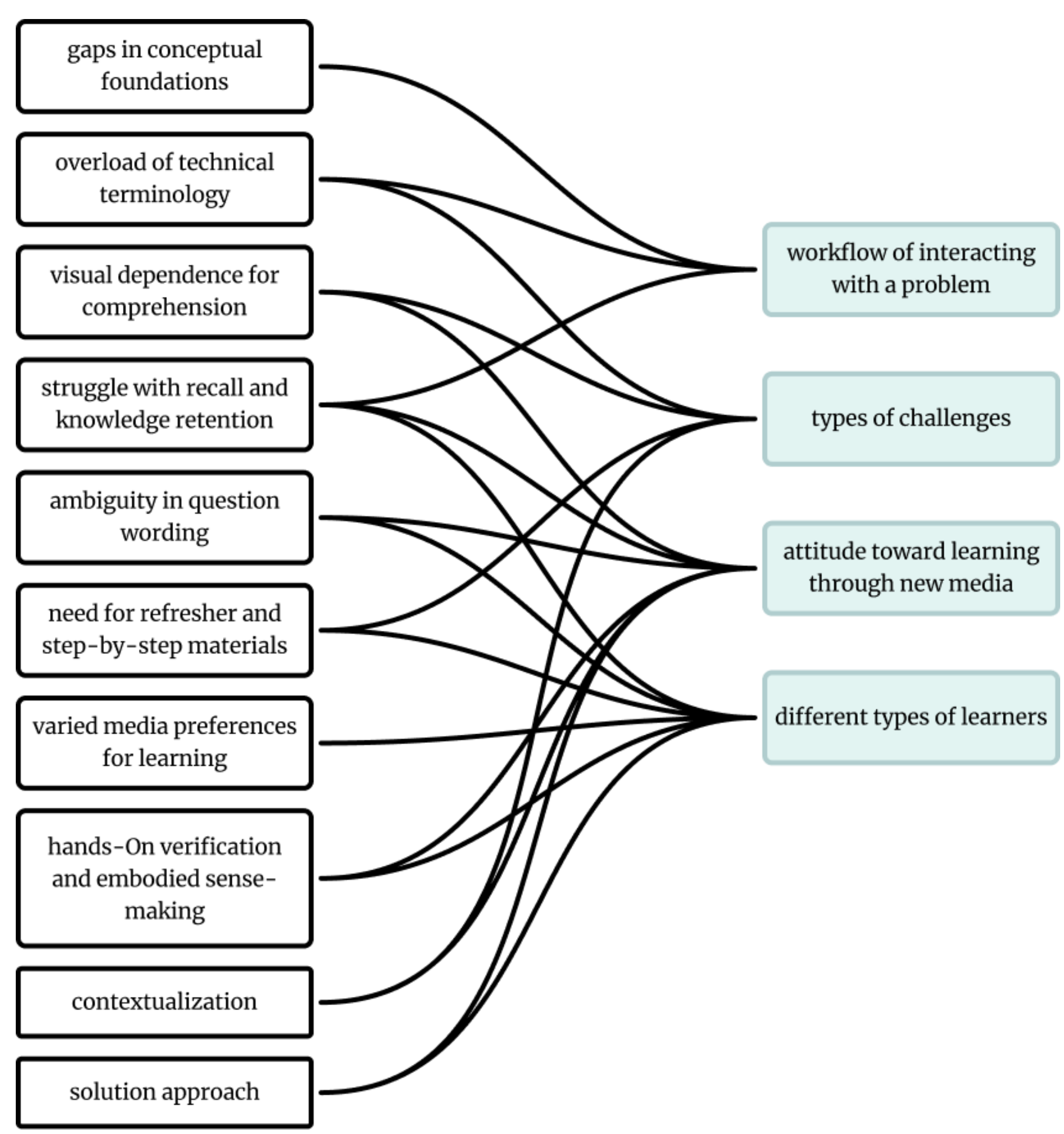

**Theme 1: Workflow of Interacting with a Fundamental Mechanics Problem**

Across instructional modalities, participants relied on standard problem-solving strategies such as carefully reading questions, sketching diagrams, and evaluating solution approaches. In the Traditional Classroom (TC) condition, problem-solving often involved extensive notetaking during lectures, an approach many participants described as cognitively demanding and reliant on memory recall. When reflecting on their understanding of mechanical concepts, participants highlighted difficulties in applying theoretical knowledge. For example:

> ***ET Participant #1 (ET 1)****: "I had difficulty understanding how the bending moment would change depending on the different external forces acting on a beam."*
>
> ***TC Participant #2 (TC 2):*** *"Remembering the exact concepts and the relations between bending moment and shear force, I was unsure about the answers [had to check multiple solutions]."*

In contrast, participants in the Mixed Reality (MR) condition described a different workflow that relied less on notetaking and more on tangible interaction, exploration, and visualization before transitioning to sketches or conceptual reasoning. This shift raises important questions: To what extent do MR and ET learners require traditional note-taking, and in what form?

> **ET 1:** *".... I could physically place the weights on the beam and see the change in the reaction, which really helped me understand the effect of the location of the point load on the reaction forces. Having known that, it was simple to draw the shear force and moment diagrams from that."*

Participants seem to engage more effectively through direct interaction with problems than through notetaking alone. Prior work using MaRginalia [22] showed that structured digital notetaking reduces cognitive load in virtual classrooms. Building on this, our results suggest that coupling immersive environments with tangible, interactive feedback and digital notetaking can better support multimodal learning. Such integration may enable new problem-solving workflows, including adaptive strategies for classroom activities, quizzes, and virtual evaluations. A MR participant noted the need for multimodal approaches as a replacement for unnecessary notetaking, which may not have any future use at all:

> ***MR 8:*** *"I would suggest both physical and digital materials would be helpful for me to understand the concepts. Because sometimes I get zoned out in class when I can't understand or visualize the topic properly. So, a visual aid [to understand and interact] like a 3D model would be really helpful."*

These findings suggest that traditional problem-solving workflows impose high cognitive load and could be redesigned using multimodal, interactive environments to support spatial reasoning, experimentation, and creativity. Designing effective VR modules should therefore prioritize interactions that allow learners to directly see, manipulate, and experiment with mechanisms, rather than simply translating traditional classroom activities into digital form.

**Theme 2: Types of Challenges**

Across all instructional modalities, participants reported difficulties recalling and applying key mechanics concepts, even though these had been covered in a prerequisite course. This suggests that teaching modules must be designed to minimize extraneous cognitive load and support knowledge transfer.

**ET 7:** *“I had trouble remembering the content since I took it a semester ago.”*

Fundamental courses establish not only terminology but also provide the conceptual base for advanced mechanical analysis. However, participants noted that the MR and ET modules, although intended to reinforce foundational theories, sometimes presented overly complex visualizations, which hindered comprehension of interconnected concepts.

**MR 1:** *“While I was going through the questions I was trying to visualize the image in my head, and some of them I just couldn't ……[for example] just the image of the beam and how are the loads are applied could help to understand the question, based on the image it is easy to understand or visualize the diagrams.”*

Several participants suggested that extended sessions combining traditional explanations with interactive visualization would improve engagement and recall. While MR and ET tools were valued for building initial understanding, participants emphasized the need for clearer, layered visualizations that progressively introduce complex relations.

**MR 2:** *“After reviewing the video and then using the Mixed Reality example, the challenges I had before with trying to visualize the example were almost gone… I didn’t recall the solution from the video, so honestly, it’s just that the concept isn’t fully clear in my mind.”*

**TC 6:** *"The learning material helped me with the basic concepts of the different types of forces and gave me some common examples of where these types of beams are used. It also helped me understand the different types of forces and solve the calculations."*

Importantly, participants did not remain passive when facing these challenges. Instead of "sitting silently," they actively asked questions and sought clarification when confused. This indicates that immersive tools must be designed with built-in communication loops that not only enable direct interaction with the system but also facilitate ongoing dialogue between learners, educators, and designers to anticipate the needs of diverse user groups. For example, an interaction designer could analyze usage histories and user footprints to generate an index of engagement or difficulties. Such an index could then be incorporated into student progress evaluations by instructors, providing a more comprehensive view of learning.

**Theme 3: Attitude Toward Learning Through New Media**

Before being introduced to MR or ET modalities, participants expressed a strong preference for effective demonstration techniques. They emphasized that visualizing how a mechanical system works could significantly improve comprehension and aid in recalling previously learned concepts. For example:

**ET 8:** *"Visualization of the things happening would help a lot. Some concepts are easier to understand through looking."*

**MR 4:** *"A refresher on these topics, as well as some discussion on how shear and moment forces are related to each other and what type of relationship they share, would be helpful.*

*Also, an explanation of what is meant by boundary conditions, as asked in the last question."*

Participants frequently noted that enjoyment and novelty in learning experiences improved their understanding. To probe further, we asked what could be added to engineering mechanics courses to make them more engaging and how they perceived MR and ET modalities.

**MR 3:** *"Fairly effective. It covered most of the topics found in the questions and reminded me of the rest."*

**MR 4:** *"The materials were quite effective, as I felt much more confident after having both."*

Interestingly, several participants distinguished between modalities: videos provided helpful conceptual explanations, while MR primarily supported visualization of mechanical systems but did not always translate directly into problem-solving success. This distinction highlights a critical design challenge: how to align visualization tools with problem-solving tasks rather than treating them as parallel aids. While nearly all participants agreed that ET and MR enhanced engagement, they also emphasized that the evaluation method strongly influenced their experience of the system. These findings suggest that immersive learning tools should not only provide visual and interactive representations but also incorporate mechanisms to surface misconceptions and track where learners encounter difficulties, enabling instructors to adapt teaching and assessment accordingly.

**Theme 4: Different Types of Learners**

Our analysis of participant interactions with MR and ET learning modalities revealed that many participants were not fully satisfied with the modality they were assigned. Some MR participants noted that greater tangibility could enhance their learning experience, while ET participants appreciated the hands-on, cause-and-effect interactions but felt constrained by the lack of visualization. We also observed trends in participants' use of VR headsets, tangible toolkits, and problem-solving strategies. Rather than classifying users as novice or expert, we examined how they perceived their assigned modalities. Based on these perceptions, we identified three learner types:

***1. Knowing Through Doing Learners:***

These learners demonstrated curiosity by actively engaging with tangible tools and asking exploratory questions. They were motivated not only by solving the immediate task but also by extending their knowledge to future applications. This aligns with ergonomic principles of active engagement and embodied interaction, which can reduce cognitive load by grounding abstract concepts in physical experience.

> **ET 8:** *"The material shown in this study was very helpful in understanding different scenarios that occur in solid mechanics, especially the beam experiment with the scale showing the different forces that occur when weight is applied at different points."*

***2. Tactile-visual Learners***

These learners valued rich visualizations and real-world examples within digital or virtual environments. They were highly interactive in VR/ET settings and expressed a desire for more advanced digital toolkits. For these learners, visualization supports spatial reasoning and complements tactile interaction, suggesting that multimodal representations are critical for their comprehension.

**MR 8:** *“A better knowledge of certain topics and applying these to a real-life scenario could help me understand certain topics better.”*

***3. Goal-oriented Learners***

These learners prioritized efficiency and outcomes over exploration. Their focus was on reaching correct answers quickly using available resources, with less interest in playful or creative engagement. For these learners, system design should minimize extraneous interactions and provide streamlined workflows, supporting accuracy and reducing time-on-task.

**TC 5:** *“I faced challenges with mostly visualizing the different types of scenarios that were given to me. After watching the video, I had a better understanding of the concept, but there are now more different ways forces have been applied, making every situation different. I faced difficulty understanding exactly why one situation behaves differently than the other.”*

## Implications and Design Guidelines

Based on participants’ experiences and feedback, the following design guidelines translate empirical findings into actionable recommendations for mixed-reality and tangible learning tools in engineering education.

1. **Integrate Mixed-Reality Visualization with Physical Interaction to Support Learning and Sustain Engagement.** Participants valued the combination of tangibility and mixed reality, reporting improved conceptual understanding and hands-on exploration. Although ET and MR conditions produced the highest quiz scores, feedback indicated that relying on a single modality could reduce engagement over time due to diminished novelty, underscoring the need for integrated multimodal designs.

2. **Design for Diverse Learning Preferences through Multiple Interaction Pathways.** Findings revealed variability in learning styles and engagement, with participants responding differently to tactile, visual, and goal-oriented approaches. Providing multiple ways to interact with content rather than a single dominant mode can better accommodate diverse learners and support effective problem-solving workflows.

3. **Use Intuitive, Multisensory Visualizations and Evolve Assessment Strategies to Reduce Cognitive Load.** Participants emphasized the importance of clear visual aids, such as familiar graphs, heatmaps, and step-by-step animations, to improve comprehension of complex mechanics concepts and reduce cognitive effort. To better capture learning with MR and ET tools, evaluation approaches should extend beyond traditional testing to include engagement and conceptual understanding, supported by adaptive feedback and instructional scaffolding.

## Conclusion

This work contributes a human-centered, comparative analysis of mixed-reality applications and physical toolkits, revealing how interaction modality shapes learner workflows, engagement, and conceptual reasoning in Engineering Mechanics. The findings highlight that multimodal instructional approaches combining hands-on interaction with clear, well-structured visualizations are particularly effective for reducing cognitive load and supporting learner needs, whereas poorly aligned or complex visualizations can hinder understanding. These results emphasize the importance of intuitive design, layered visual representations, and alignment with learners' problem-solving processes. While this study involved a small sample and focused on a single Engineering Mechanics topic outside a full course context, future work should examine broader implementation across topics and instructional settings. Ultimately, future engineering learning environments should move beyond standalone tools toward integrated, learner-centered instructional ecosystems that combine physical interaction, immersive visualization, and adaptive assessment to support sustained engagement and conceptual understanding.

## Declaration of Conflicting Interests

The authors declare no potential conflicts of interest with respect to the research, authorship, or publication of this article.

## Funding

This research was supported by the Research Experiences for Undergraduates (REU) Program of the University of Texas at Arlington College of Engineering.

## Key Points

1. Mixed Reality and physical toolkits improve visualization and engagement in engineering education.
2. Learners prefer multimodal approaches combining hands-on interaction with clear visual aids.
3. Design strategies should prioritize intuitive, interactive workflows to support diverse learner types.